\documentclass[seceq]{ptptex}

\usepackage{graphicx}

\newcommand{\bra}{\langle}
\newcommand{\ket}{\rangle}
\newcommand{\R}{{\mathbb R}}
\newcommand{\C}{{\mathbb C}}
\newcommand{\Tr}{\mbox{Tr} \,}

\title{
A method for systematic construction of Bell-like inequalities
\\
and a proposal of a new type of test
}

\author{
Tomohiro \textsc{Isobe} 
and Shogo \textsc{Tanimura}\footnote{E-mail: tanimura@i.kyoto-u.ac.jp}
}

\inst{
Department of Applied Mathematics and Physics,
Graduate School of Informatics,
\\
Kyoto University, Kyoto 606-8501, Japan
}

\abst{
The Bell-Clauser-Horne-Shimony-Holt (BCHSH) inequality,
which is proven in the context of the local hidden variable theory,
has been used as a test 
to reveal failure of the hidden variable theory 
and to prove validity of the quantum theory.
We note that violation of the BCHSH inequality is caused by
noncommutativity of quantum observables
and find a systematic method for constructing generalizations of 
the BCHSH inequality.
This method is applied for inventing a new quantity
which is defined in a two-qubit system and
satisfies a new type of inequality.
This provides a fair test of the quantum theory.
Remaining problems are also discussed.
}

\PTPindex{060, 061}  

\begin{document}

\maketitle

\section{Introduction}
{}From the early time in the history of quantum mechanics,
doubts about validity or completeness of quantum mechanics have been raised repeatedly.
For instance, Einstein~\cite{EPR1935} and Bohr~\cite{Bohr1935}
devoted themselves to enthusiastic debates.
De Broglie proposed the pilot wave theory as an alternative to quantum mechanics.
Bohm~\cite{Bohm1952} and other people developed 
de Broglie's idea as the hidden variable theory.
Bell~\cite{Bell1964} formulated his famous inequality 
in the context of the hidden variable theory and
proposed a possible test to compare prediction of quantum mechanics 
with predictions of other alternative theories.
Since his proposal, many experiments have been 
performed~\cite{CH1978, Aspect1981}
and they support validity of quantum mechanics with increasing accuracy~\cite{Sakai2006}.
Many people~\cite{Froissart1981, Peres1999, Collins2004}
have proposed generalizations of the Bell inequality.
So, today there seems no room for putting a doubt on validity of quantum mechanics.

However, in this paper we attempt to propose a new test for checking validity
of quantum mechanics from a different viewpoint.
At least, we give a new explanation on the Bell inequality.
This explanation gives an insight for understanding implication of quantum mechanics
and suggests a way for producing various kinds of paradoxical inequalities
which are not equivalent to the original Bell inequality.

The hidden variable theory consists of the following assumptions:
(1) The state of a physical system is specified 
not only by a quantum state $ \psi $,
but also by a variable $ \lambda $ 
or a set of variables $ \lambda = ( \lambda_1, \lambda_2, \cdots , \lambda_n ) $
which we do not know.
(2) Any physical observable $ A $ is a function of $ \psi $ and $ \lambda $.
Once the state $ \psi $ and the value of the hidden variable $ \lambda $ are specified,
the value $ A ( \psi, \lambda ) \in \R $ of the observable is uniquely determined.
(3) The variable $ \lambda $ obeys a probability distribution $ P( \psi, \lambda ) $,
which is nonnegative, $ P \ge 0 $, and normalized, $ \int P d \lambda = 1 $.
(4) Additionally, in the local hidden variable theory,
the value or the distribution of the hidden variable cannot be
influenced by superluminous signals.

In the scheme of the hidden variable theory,
the expectation value of the observable $ A $ is calculated as
\begin{equation}
	\bra A \ket = \int A ( \psi, \lambda ) P( \psi, \lambda ) \, d \lambda
	\label{hidden variable}
\end{equation}
with the probability distribution $ P( \psi, \lambda ) $ 
of the hidden variable $ \lambda $.
So, a question arises; 
can the hidden variable theory mimic the quantum theory completely?
In other words,
does the probability distribution of the hidden variable 
which reproduces the predictions of the quantum theory for any observables
mathematically exist?

Bell proved an inequality which bounds the expectation value of a certain observable
in the scheme of the local hidden variable theory.
Clauser, Horne, Shimony and Holt~\cite{CHSH1969, CH1974}
reformulated Bell's inequality
in a form more suitable for experimental tests.
It tells that the expectation value of a quantity 
$ S = A_1 B_1 + A_1 B_2 + A_2 B_1 - A_2 B_2 $
must be in the range
\begin{equation}
	-2 \: \le \: \bra S \ket \: \le \: 2
	\qquad
	\mbox{: hidden variable theory}
	\label{BCHSH inequality},
\end{equation}
if the local hidden variable theory is correct.
We call (\ref{BCHSH inequality}) the BCHSH inequality
abbreviating the names of the authors; Bell, Clauser, Horne, Shimony and Holt.
On the other hand, the quantum theory predicts that
\begin{equation}
	-2 \sqrt{2} \: \le \: \bra S \ket \: \le \: 2 \sqrt{2}
	\qquad
	\mbox{: quantum theory}.
	\label{quantum inequality}
\end{equation}
If experiments yield the value of $ \bra S \ket $ in the range
$ -2 \sqrt{2} \le \bra S \ket < -2 $
or in
$ 2 < \bra S \ket \le 2 \sqrt{2} $,
the BCHSH inequality is violated and
we can conclude that the hidden variable theory is wrong and the quantum theory is correct.
After proposal of the BCHSH inequality,
many experiments have been performed~\cite{CH1978, Aspect1981, Aspect1982, Sakai2006} 
and they revealed violation of the BCHSH inequality.
So, there is no doubt of failure of the hidden variable theory.

However, the meaning of the quantity 
$ S = A_1 B_1 + A_1 B_2 + A_2 B_1 - A_2 B_2 $
seems obscure.
A lot of generalizations of the BCHSH inequality have been proposed
by other researchers~\cite{Froissart1981, Peres1999, Collins2004}
and a systematic method for generalization also has been given by 
Avis, Moriyama, and Owari~\cite{Avis2009},
who used methods of the operations research.
But it is still desirable to construct Bell-like inequalities
with understanding of the physical principle which enables the construction.
In addition, it is noted that the BCHSH inequality is not a fair test
for the hidden variable theory in a sense explained below.

We classify types of inequalities which examine validity
of the hidden variable theory and the quantum theory.
In general, for a physical quantity $ T $,
each theory predicts that the expectation value of $ T $ falls in some range as
\begin{eqnarray}
&&	a \: \le \: \bra T \ket \: \le \: b
	\qquad
	\mbox{: hidden variable theory},
	\label{general BCHSH inequality}
	\\
&&	c \: \le \: \bra T \ket \: \le \: d
	\qquad
	\mbox{: quantum theory}.
	\label{general quantum inequality}
\end{eqnarray}
Then by measuring the experimental value
we can judge validity of the two theory.
We classify tests into four types:
\begin{eqnarray}
&&	\mbox{type 1: } \,
	c < a < b < d
	\nonumber \\
&&	\mbox{type 2: } \,
	c < a < d < b, \quad \mbox{or} \quad a < c < b < d
	\nonumber \\
&&	\mbox{type 3: } \,
	a < c < d < b
	\nonumber \\
&&	\mbox{type 4: } \,
	c < d < a < b, \quad \mbox{or} \quad a < b < c < d.
	\label{classification}
\end{eqnarray}
\begin{figure}[tb]
\begin{center}
\scalebox{0.67}{
\includegraphics{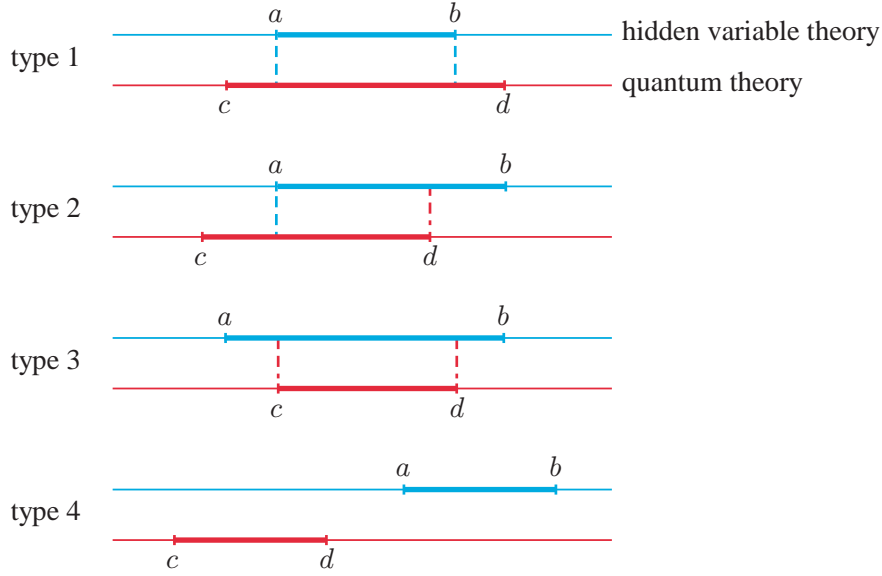}
}
\end{center}
\vspace{-2mm}
\caption{\label{FIG1}Classification of tests of
the hidden variable theory and the quantum theory.}
\end{figure}

According to this scheme,
the BCHSH inequality belongs to the type 1,
where the range of prediction of the hidden variable theory is included 
in the range of the quantum theory.
So, in any experiment,
it cannot happen that only the hidden variable theory is correct
and the quantum theory is wrong.
However, if we have a test of the type 2 with $ c < a < d < b $
and if we get the experimental value in $  d < \bra T \ket \le b $,
we should conclude that 
the hidden variable theory is correct and the quantum theory is wrong.
On the other hand,
the Kochen-Specker theorem~\cite{Kochen1967}
and the 
Greenberger-Horne-Shimony-Zeilinger test~\cite{Mermin1990, GHSZ1990, Mermin1993}
belong to the type 4,
where the range of the hidden variable theory and
the range of the quantum theory are completely disjoint.
Of course, we do not expect that any experiments invalidate the quantum theory
in the real world.
But it is desirable for strengthening validity of the quantum theory
to have a test which can reveal 
even failure of the quantum theory and success of the hidden variable theory.
Passing such a severe test like type 2 or type 4, 
the quantum theory will become more persuasive and reliable.

In this paper we explain 
several mathematical reasons of violation of the BCHSH inequality.
We also give a method for making systematic generalizations 
of the BCHSH inequality;
this method is a main result of this paper.
As a product of the main result, we invent a quantity
\begin{eqnarray}
	T 
&=&
	A_1 B_3 
	+ A_1 B_6 
	+ A_2 B_3 
	- A_2 B_6 
	\nonumber \\ &&
	+ A_3 B_2 
	+ A_3 B_5 
	+ A_1 B_2 
	- A_1 B_5 
	\nonumber \\ &&
	+ A_2 B_1 
	+ A_2 B_4 
	+ A_3 B_1 
	- A_3 B_4,
	\label{T intro}
\end{eqnarray}
where the observables 
$ A_i $ $(i=1,2,3) $ and $ B_j $ $(j=1,\cdots,6) $ take their values in $ \{ 1, -1 \} $
and $ A_i $ commutes with $ B_j $.
We will show that the two theory predict the range of the expectation value as
\begin{eqnarray}
&&	-6 \; \le \; \bra T \ket \; \le \; 6
	\qquad \qquad
	\mbox{: hidden variable theory}, 
	\nonumber \\
&&	-6 \sqrt{2} \; \le \; \bra T \ket \; \le \; 2 \sqrt{2}
	\quad 
	\mbox{: quantum theory}.
\end{eqnarray}
Hence this set of inequalities belongs to the type 2,
which offers a severer and fairer test 
for comparing the quantum theory and the hidden variable theory
than the conventional BCHSH inequality.

\section{Bell-Clauser-Horne-Shimony-Holt inequality}
In this section we present a brief review of the BCHSH inequality.
The constituents of the BCHSH inequality are four observables
$ A_1 $, $ A_2 $, $ B_1 $ and $ B_2 $.
Each observable takes $ +1 $ or $ -1 $ as its value.
It is assumed that $ A_i $ and $ B_j $ $ (i,j=1,2) $
are simultaneously measurable.
However,
$ A_1 $ and $ A_2 $ are not necessarily simultaneously measurable.
$ B_1 $ and $ B_2 $ are not either.
The quantity $ S $ is defined as
\begin{eqnarray}
	S 
	&=& A_1 B_1 + A_1 B_2 + A_2 B_1 - A_2 B_2
	\nonumber \\
	&=& A_1 ( B_1 + B_2 ) + A_2 ( B_1 - B_2 ).
	\label{S}
\end{eqnarray}

The above formulation is interpreted as follows.
Suppose we have a pair of spin-half particles or a pair of photons.
The two particles are labeled with $ A $ and $ B $, respectively.
The observable $ A_i $
is interpreted as a spin component of the spin-half particle
or a polarization of the photon.
The index $ i =1,2 $ specifies 
the direction of the polarization detector.
The observable $ B_j $
is interpreted as a spin component of the other particle
or a polarization of the other photon.
Two detectors acting on the two particles $ A $ and $ B $ 
are spatially separated,
and hence, an event observed at one detector 
cannot make influence on an event observed at the other detector.
This separation justifies defining the value of $ A_i B_j $ by 
a product of observed values of $ A_i $ and $ B_j $.
So, by varying the directions of the detectors and 
by generating pairs of particles repeatedly,
we accumulate data for the combined observables, 
$ (A_1, B_1) $,
$ (A_1, B_2) $,
$ (A_2, B_1) $ and
$ (A_2, B_2) $.
By making product of the measured values and 
by taking their average and adding them up,
we get the average of $ S $,
\begin{eqnarray}
	\bra S \ket
	&=& 
	\bra A_1 B_1 \ket 
	+ \bra A_1 B_2 \ket
	+ \bra A_2 B_1 \ket
	- \bra A_2 B_2 \ket.
	\label{average of S}
\end{eqnarray}
The hidden variable theory and the quantum theory
give different predictions on the range of $ \bra S \ket $
as seen below.

In the context of the hidden variable theory,
the values
$ A_i ( \psi, \lambda ) $, 
$ B_j ( \psi, \lambda ) 
\in \{ +1, -1 \} $
are determined depending on the quantum state $ \psi $ 
and the hidden variable $ \lambda $ of the system.
When $ B_1 + B_2 = \pm 1 \pm 1= \pm 2 $, we have $ B_1 - B_2 = 0 $.
When $ B_1 + B_2 = \pm 1 \mp 1=0 $, we have $ B_1 - B_2 = \pm 2 $.
So, one of $ (B_1+B_2) $ or $(B_1-B_2) $ is always 0 and the other is $ \pm 2 $.
The coefficients $ A_1, A_2 $ are also $ +1 $ or $ -1 $.
Hence, possible values of the quantity
$ S = A_1 ( B_1 + B_2 ) + A_2 ( B_1 - B_2 ) $
are $ \pm 2 $.
Since the probability distribution is assumed to be nonnegative and normalized,
the average 
\begin{eqnarray}
	\bra S \ket 
&=&
	\int 
	( A_1 B_1 + A_1 B_2 + A_2 B_1 - A_2 B_2 )
	\, P ( \psi, \lambda ) \, d \lambda
	\label{local calculation}
\end{eqnarray}
is in $ -2 \le \bra S \ket \le 2 $.
This proves the BCHSH inequality (\ref{BCHSH inequality}).

Let us turn to the quantum theory.
In the quantum theory, 
$ A_i $ and $ B_j $ are not functions but operators or matrices.
A typical choice for them is 
\begin{eqnarray}
	A_1 &=& \sigma_z \otimes I,
	\label{A_1}
	\\
	A_2 &=& (\sigma_z \cos 2 \theta + \sigma_x \sin 2 \theta) \otimes I,
	\\
	B_1 &=& I \otimes (\sigma_z \cos \theta + \sigma_x \sin \theta),
	\\
	B_2 &=& I \otimes (\sigma_z \cos \theta - \sigma_x \sin \theta).
	\label{B_2}
\end{eqnarray}
These are operators acting on the two-qubit Hilbert space $ \C^2 \otimes \C^2 $
and $ I $ is the two-dimensional identity matrix.
The parameter $ \theta $ is adjustable 
for specifying directions of the detectors.
By substituting the Pauli matrices,
we get the matrix representation for $ S $,
\begin{eqnarray}
	S_\theta
	& = & A_1 B_1 + A_1 B_2 + A_2 B_1 - A_2 B_2 
	\nonumber \\
	& = &
	2 \cos \theta 
	\begin{pmatrix}
	1 & 0 & 0 & 0 \\ 
	0 & -1 & 0 & 0 \\ 
	0 & 0 & -1 & 0 \\ 
	0 & 0 & 0 & 1 
	\end{pmatrix}
	\nonumber \\
	&& 
	+ 2 \cos 2 \theta \sin \theta 
	\begin{pmatrix}
	0 & 1 & 0 & 0 \\ 
	1 & 0 & 0 & 0 \\ 
	0 & 0 & 0 & -1 \\ 
	0 & 0 & -1 & 0 
	\end{pmatrix}
	+ 2 \sin 2 \theta \sin \theta 
	\begin{pmatrix}
	0 & 0 & 0 & 1 \\ 
	0 & 0 & 1 & 0 \\ 
	0 & 1 & 0 & 0 \\ 
	1 & 0 & 0 & 0 
	\end{pmatrix}.
	\label{S theta}
\end{eqnarray}
The eigenvalues of $ S_\theta $ are
\begin{eqnarray}
&&	\{ s_1 (\theta), s_2 (\theta), s_3 (\theta), s_4 (\theta) \}
	\nonumber \\
&&
	=
	\Big\{
	 2 \cos 2 \theta, \,
	-2 \cos 2 \theta, \,
	 2 \sqrt{1 + \sin^2 2 \theta}, \,
	-2 \sqrt{1 + \sin^2 2 \theta} \,
	\Big\}.
\end{eqnarray}
\begin{figure}[bt]
\begin{center}
\scalebox{0.50}{
\includegraphics{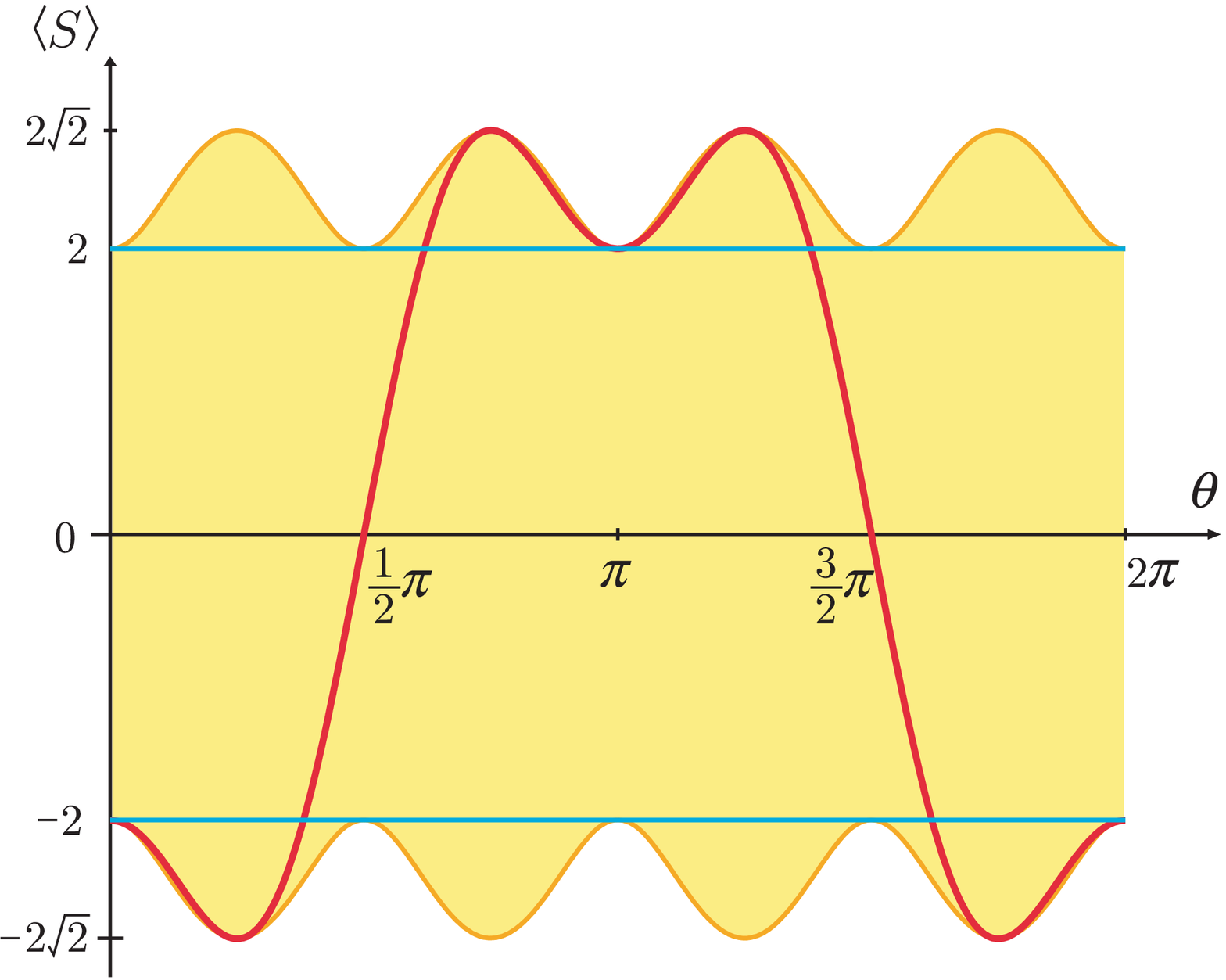}
}
\end{center}
\vspace{-2mm}
\caption{\label{FIG2}A test of 
the ordinary Bell-Clauser-Horne-Shimony-Holt inequality.
The hidden variable theory predicts that $ -2 \le \bra S \ket \le 2 $.
The quantum theory predicts that $ \bra S \ket $ is in the yellow band
$ B_S $ defined in (\ref{B_S}).
The red curve is the expectation value (\ref{singlet value S})
associated with the singlet state.}
\end{figure}
For a general state we get an expectation value
$ \bra S_\theta \ket = \sum_{i=1}^4 w_i s_i (\theta) $,
which is a convex combination of $ \{ s_i (\theta) \}_{i=1,\cdots,4} $.
Then we introduce sets of expectation values as
\begin{eqnarray}
	Q_S ( \theta ) 
	& :=&
	\Big\{ 
	\sum_{i=1}^4 w_i s_i (\theta) \, \Big| \, 
	w_i \in \R, \, 0 \le w_i \le 1, \, \sum_{i=1}^4 w_i = 1 \,
	\Big\}
	\nonumber \\
	&=& 
	\Big[ -2 \sqrt{1 + \sin^2 2 \theta}, 
	\; 2 \sqrt{1 + \sin^2 2 \theta} \, \Big] \subset \R,
\\
	Q_S 
	& := & \bigcup_{ 0 \le \theta \le 2 \pi} Q_S ( \theta ) 
	= \big[ -2 \sqrt{2}, \; 2 \sqrt{2} \, \big]
	\subset \R,
\\
	B_S 
	& := & 
	\big\{
	( \theta, s ) \, \big| \, 
	0 \le \theta \le 2 \pi, \, s \in Q_S (\theta) 
	\big\}
	\subset \R^2.
	\label{B_S}
\end{eqnarray}
Hence the range $ Q_S $ implies that
$ -2 \sqrt{2} \le \bra S \ket \le 2 \sqrt{2} $ in the quantum theory.
This proves the inequality (\ref{quantum inequality}).

We proved that the range of the expectation value 
allowed by the hidden variable theory is
\begin{eqnarray}
	H_S := [ -2, \, 2 \, ] \subset \R.
\end{eqnarray}
Then it holds that $ H_S \subset Q_S (\theta) $ for any $ \theta $.
So, the quantity $ S $ provides only the type 1 test.
The band $ B_S $ formed by values allowed by the quantum theory
is shown in the figure \ref{FIG2} as the painted domain.

In particular, if we take the spin singlet state
\begin{eqnarray}
	\psi
	&=&
	\psi_1 + \psi_2
	=
	| \! \uparrow \downarrow \ket + | \! \downarrow \uparrow \ket 
	\nonumber \\
	&=&
	\frac{1}{\sqrt{2}}
	\begin{pmatrix} 1 \\ 0 \end{pmatrix}
	\otimes
	\begin{pmatrix} 0 \\ 1 \end{pmatrix}
	-
	\frac{1}{\sqrt{2}}
	\begin{pmatrix} 0 \\ 1 \end{pmatrix}
	\otimes
	\begin{pmatrix} 1 \\ 0 \end{pmatrix}
	=
	\frac{1}{\sqrt{2}}
	\begin{pmatrix} 0 \\ 1 \\ -1 \\ 0 \end{pmatrix},
	\label{singlet}
\end{eqnarray}
the expectation value becomes
\begin{eqnarray}
	\bra \psi | S_\theta | \psi \ket
	& = &
	- 2 \cos \theta 
	- 2 \sin 2 \theta \sin \theta.
	\label{singlet value S}
\end{eqnarray}
At $ \theta = \pm \frac{\pi}{4} $,
we get $ \bra \psi | S_\theta | \psi \ket = -2 \sqrt{2} $.
At $ \theta = \pm \frac{3}{4} \pi $,
we get $ \bra \psi | S_\theta | \psi \ket = 2 \sqrt{2} $.
Thus the maximum violation of the BCHSH inequality is attained at these angles.

Here we explain implication of locality.
In the context of the hidden variable theory,
locality or nonlocality is formulated as follows.
Locality requests that the probability distribution 
is independent of the directions of the detectors,
which are placed far 
from each other and from the source of the particle pair.
Namely, in the local theory we calculate the average with the formula
\begin{equation}
	\bra A_i B_j \ket 
	= \int A_i ( \psi, \lambda ) B_j ( \psi, \lambda ) 
	P ( \psi, \lambda ) \, d \lambda
	\quad
	\mbox{: local hidden variable theory}.
	\label{local}
\end{equation}
On the other hand, the nonlocal hidden variable theory allows that
the probability distribution depends on the directions 
of the far separated detectors.
Hence, the above formula is replaced by
\begin{equation}
	\bra A_i B_j \ket 
	= \int A_i ( \psi, \lambda ) B_j ( \psi, \lambda ) 
	P_{ij} ( \psi, \lambda ) \, d \lambda
	\quad
	\mbox{: nonlocal hidden variable theory}.
	\label{nonlocal}
\end{equation}
The calculation (\ref{local calculation}) is based 
on the local hidden variable theory,
not on the nonlocal theory.
It is to be emphasized that the BCHSH inequality is proved 
in the context of the local hidden variable theory.

In the context of the quantum theory, 
we adopt the conventional interpretation which tells that
locality implies commutativity of observables
separated by a spacelike distance.
The operators $ A_i \, (i=1,2) $ act on the Hilbert space of the particle $ A $
while
the operators $ B_j \, (j=1,2) $ act on the Hilbert space of the particle $ B $.
So, they are acting on different spaces, and hence they commute.

It should be mentioned that 
commutativity is not a necessary condition for locality.
Deutsch~\cite{Deutsch2004} constructed 
a model in which spacelike-separated observables do not commute
and showed that no contradiction arises in his model.
Hence, identification of locality and commutativity 
should not be accepted without question.

\section{Why is the BCHSH inequality violated?}
In this section we give several explanations on violation of the BCHSH inequality.
Here we consider in the context of the quantum theory.

One of the interpretations of the violation is interference effect.
If the system is in a mixed state described by the density matrix
\begin{eqnarray}
	\rho
&=&
	| \psi_1 \ket \bra \psi_1 | +
	| \psi_2 \ket \bra \psi_2 |
	\nonumber \\
&=&
	\frac{1}{2}
	\begin{pmatrix} 1 & 0 \\ 0 & 0 \end{pmatrix}
	\otimes
	\begin{pmatrix} 0 & 0 \\ 0 & 1 \end{pmatrix}
	+
	\frac{1}{2}
	\begin{pmatrix} 0 & 0 \\ 0 & 1 \end{pmatrix}
	\otimes
	\begin{pmatrix} 1 & 0 \\ 0 & 0 \end{pmatrix}
	=
	\frac{1}{2}
	\begin{pmatrix}
	0 & 0 & 0 & 0 \\
	0 & 1 & 0 & 0 \\
	0 & 0 & 1 & 0 \\
	0 & 0 & 0 & 0 
	\end{pmatrix},
\end{eqnarray}
the expectation value becomes
\begin{eqnarray}
	\Tr ( S_\theta \, \rho )
	= \bra \psi_1 | S_\theta | \psi_1 \ket
	+ \bra \psi_2 | S_\theta | \psi_2 \ket
	=
	- 2 \cos \theta
\end{eqnarray}
and it stays in the range $ -2 \le \Tr ( S_\theta \, \rho ) \le 2 $
being consistent with the BCSHS inequality.
Instead of the mixed state, 
if we substitute the pure state $ \rho = | \psi \ket \bra \psi | $
with (\ref{singlet}),
the off-diagonal elements of $ S_\theta $ in (\ref{S theta})
contribute to the expectation value as
\begin{eqnarray}
	\bra \psi | S_\theta | \psi \ket
&=&
	  \bra \psi_1 | S_\theta | \psi_1 \ket
	+ \bra \psi_2 | S_\theta | \psi_2 \ket
	+ \bra \psi_1 | S_\theta | \psi_2 \ket
	+ \bra \psi_2 | S_\theta | \psi_1 \ket
	\nonumber \\
&=&
	- 2 \cos \theta 
	- 2 \sin 2 \theta \sin \theta.
\end{eqnarray}
The cross terms
$ \bra \psi_1 | S_\theta | \psi_2 \ket
+ \bra \psi_2 | S_\theta | \psi_1 \ket $
represent interference effect,
which causes the violation of the BCHSH inequality.
The interference of the two terms in the superposed state
$ \psi = | \!\! \uparrow \downarrow \ket + | \!\! \downarrow \uparrow \ket $
is also called entanglement effect.
This kind of explanation for the violation of the BCHSH inequality
can be found in recent textbooks~\cite{Shimizu2003}.

\section{Method for systematic construction of Bell-like inequalities}
Here we give another explanation for the violation of the BCHSH inequality,
which gives a hint for generalizing the inequality.
This explanation is based on noncommutativity of quantum observables.
The eigenvalues of
$ B_1 = \sigma_z \cos \theta + \sigma_x \sin \theta $
are $ \{ 1, -1 \} $.
The eigenvalues of 
$ B_2 = \sigma_z \cos \theta - \sigma_x \sin \theta $
are $ \{ 1, -1 \} $, too.
However, the eigenvalues of $ B_1 + B_2 $
are not $ \{ 2, 0, -2 \} $.
Actually, the eigenvalues of
$ B_1 + B_2 = \sigma_z \, 2 \cos \theta $
are $ \{ 2 \cos \theta, -2 \cos \theta \} $,
which become $ \{ \sqrt{2}, - \sqrt{2} \} $
at $ \theta = \pm \frac{\pi}{4} $ or $ \pm \frac{3}{4} \pi $ particularly.
This trick is written in the typical form as
\begin{eqnarray}
	\frac{1}{\sqrt{2}} ( \sigma_z + \sigma_x ) +
	\frac{1}{\sqrt{2}} ( \sigma_z - \sigma_x ) =
	\sqrt{2} \, \sigma_z.
	\label{trick}
\end{eqnarray}
While each term $ \frac{1}{\sqrt{2}} ( \sigma_z \pm \sigma_x ) $ 
in the left hand side has the spectrum $ \{ 1, -1 \} $,
the right term $ \sqrt{2} \, \sigma_z $
has the spectrum $ \{ \sqrt{2}, - \sqrt{2} \} $.
Symbolically, we can say that
{\em 1 + 1 is not 2 but $ \sqrt{2} $ in the quantum theory.}
This example tells that
{\it the eigenvalues of a sum of noncommutative operators 
are not equal to the sum of eigenvalues of the respective operators in general}.
This is an elementary fact of linear algebra.
The sum or the product of eigenvalues of operators $ B_1 $ and $ B_2 $
is equal to
eigenvalues of $ B_1 + B_2 $ or $ B_1 B_2 $
only in their simultaneous eigenvector.
Namely, the proposition
\begin{eqnarray}
	B_1 \phi_1 = b_1 \phi_1, \:
	B_2 \phi_2 = b_2 \phi_2
	\, \Rightarrow \,
	( B_1 + B_2 ) \phi = ( b_1 + b_2 ) \phi, \:
	( B_1 B_2 ) \phi = ( b_1 b_2 ) \phi
	\quad
	\label{naive}
\end{eqnarray}
holds only when the state vectors $ \phi_1 $, $ \phi_2 $ and $ \phi $
belong to the common eigenspace of $ B_1 $ and $ B_2 $.
On the other hand, if $ B_1 $ and $ B_2 $ are noncommutative,
there is a state vector which is not decomposable
into the common eigenspaces of $ B_1 $ and $ B_2 $.
Then the naive calculation like (\ref{naive}) does not hold.

The hidden variable theory assumes that 
the observables $ A_i $ and $ B_j $ have some values 
$ A_i (\lambda) $ and $ B_j (\lambda) $ at any time 
even when the observables are not measured\footnote{
In the following, we do not write
dependence on $ \psi $ of 
$ A_i (\psi, \lambda) $ and $ P (\psi, \lambda) $
explicitly.
}.
It also assumes that the values obey the ordinary arithmetic rule.
The reasoning based on these assumptions leads to
the BCHSH inequality (\ref{BCHSH inequality}).
But, in the quantum theory, 
values cannot be assigned to the noncommuting observables simultaneously
and
the naive arithmetic rule is not applicable to their values.
Hence the BCHSH inequality is violated.

This kind of reasoning reveals the trick for making the BCHSH quantity $ S $.
We begin with 
\begin{eqnarray}
	S 
	=
	\sqrt{2} \, 
	( \sigma_z \otimes \sigma_z 
	+ \sigma_x \otimes \sigma_x ).
	\label{begin}
\end{eqnarray}
Note that the spectra of
$ \sigma_z \otimes \sigma_z $ and $ \sigma_x \otimes \sigma_x $
are both $ \{ 1, -1 \} $.
Since 
$ \sigma_z \otimes \sigma_z $ and $ \sigma_x \otimes \sigma_x $ commute,
the naive arithmetic rule is applicable to them and 
the spectrum of $ S $ should be a subset of 
$ \{ 2 \sqrt{2}, \, 0, \, -2 \sqrt{2} \} $;
actually the spectrum of $ S $ is $ \{ 2 \sqrt{2}, \, -2 \sqrt{2} \} $.
Then by applying the trick (\ref{trick}) for rewriting $ S $ we get
\begin{eqnarray}
	S 
&=&
	\sqrt{2} \, 
	( \sigma_z \otimes \sigma_z 
	+ \sigma_x \otimes \sigma_x )
	\nonumber \\
&=&
	\sqrt{2} \Big[
	   \sigma_z \otimes \frac{1}{2} \{ ( \sigma_z + \sigma_x ) + ( \sigma_z - \sigma_x ) \}
	+ \sigma_x \otimes \frac{1}{2} \{ ( \sigma_z + \sigma_x ) - ( \sigma_z - \sigma_x ) \}
	\Big]
	\nonumber \\
&=&
	  \sigma_z \otimes \frac{1}{\sqrt{2}} ( \sigma_z + \sigma_x ) 
	+ \sigma_z \otimes \frac{1}{\sqrt{2}} ( \sigma_z - \sigma_x ) 
	\nonumber \\ && 
	+ \sigma_x \otimes \frac{1}{\sqrt{2}} ( \sigma_z + \sigma_x ) 
	- \sigma_x \otimes \frac{1}{\sqrt{2}} ( \sigma_z - \sigma_x ) 
	\nonumber \\
&=&	A_1 B_1 + A_1 B_2 + A_2 B_1 - A_2 B_2,
	\label{how to make it}
\end{eqnarray}
which is the quantity maximally violating the BCHSH inequality.
By introducing the adjustable parameter $ \theta $
we get the quantity $ S_\theta $ expressed 
in terms of the observables (\ref{A_1})-(\ref{B_2})

Actually, Bell~\cite{Bell1966, Mermin1993} himself noticed that
values of noncommuting observables 
do not obey the naive additive law (\ref{naive})
but he did not utilize this property to derive his inequality.
Seevinck and Uffink~\cite{Seevinck2007} argued that
noncommutativity is related to violation of the original BCHSH inequality
but they did not consider a method 
for generating variations of the BCHSH inequality.

\section{New inequality}
Here we build a new quantity $ T $
which satisfies a new type of the Bell-like inequality.
We begin with the quantity
\begin{equation}
	T = 2 \sqrt{2} \, (
	\sigma_x \otimes \sigma_x +
	\sigma_y \otimes \sigma_y +
	\sigma_z \otimes \sigma_z ).
	\label{starting T}
\end{equation}
The three terms $ \sigma_a \otimes \sigma_a $ $ ( a = x,y,z ) $
are mutually commutative
and their spectra are $ \{ 1, -1 \} $.
Therefore, the spectrum of $ T $ should be a subset of 
$ \{ 6 \sqrt{2}, \, 2 \sqrt{2}, \, -2 \sqrt{2}, \, -6 \sqrt{2} \} $.
The matrix representation of $ T $ is calculated as
\begin{eqnarray}
	\frac{1}{2 \sqrt{2}} \, T 
&=&
	\begin{pmatrix} 0 & 1 \\ 1 & 0 \end{pmatrix} 
	\otimes
	\begin{pmatrix} 0 & 1 \\ 1 & 0 \end{pmatrix}
	+
	\begin{pmatrix} 0 & -i \\ i & 0 \end{pmatrix} 
	\otimes
	\begin{pmatrix} 0 & -i \\ i & 0 \end{pmatrix} 
	+
	\begin{pmatrix} 1 & 0 \\ 0 & -1 \end{pmatrix} 
	\otimes
	\begin{pmatrix} 1 & 0 \\ 0 & -1 \end{pmatrix} 
	\nonumber \\
&=&
	\begin{pmatrix}
	0 & 0 & 0 & 1 \\
	0 & 0 & 1 & 0 \\
	0 & 1 & 0 & 0 \\
	1 & 0 & 0 & 0 
	\end{pmatrix} 
	+
	\begin{pmatrix}
	0 & 0 & 0 &-1 \\
	0 & 0 & 1 & 0 \\
	0 & 1 & 0 & 0 \\
	-1& 0 & 0 & 0 
	\end{pmatrix} 
	+
	\begin{pmatrix}
	1 & 0 & 0 & 0 \\
	0 &-1 & 0 & 0 \\
	0 & 0 &-1 & 0 \\
	0 & 0 & 0 & 1 
	\end{pmatrix} 
	\nonumber \\
&=&
	\begin{pmatrix}
	1 & 0 & 0 & 0 \\
	0 &-1 & 2 & 0 \\
	0 & 2 &-1 & 0 \\
	0 & 0 & 0 & 1 
	\end{pmatrix}.
	\label{starting T matrix}
\end{eqnarray}
It is easily seen that the eigenvalues of $ T $
are $ 2 \sqrt{2} $ (three-fold degeneracy) and $ -6 \sqrt{2} $ (no degeneracy);
the corresponding eigenvectors are
\begin{eqnarray}
	2 \sqrt{2} : \:
	a
	\begin{pmatrix}
	1 \\ 0 \\ 0 \\ 0
	\end{pmatrix}
	+b
	\begin{pmatrix}
	0 \\ 1 \\ 1 \\ 0 
	\end{pmatrix}
	+c
	\begin{pmatrix}
	0 \\ 0 \\ 0 \\ 1 
	\end{pmatrix},
	\qquad
	-6 \sqrt{2} : 
	\begin{pmatrix}
	0 \\ 1 \\ -1 \\ 0
	\end{pmatrix}.
\end{eqnarray}
In other words,
the states with the eigenvalue $ T = 2 \sqrt{2} $
are the triplet spin state, while 
the state with the eigenvalue $ T = -6 \sqrt{2} $
is the singlet spin state.
Hence the quantum theory predicts the bound
\begin{equation}
	-6 \sqrt{2} \; \le \; \bra T \ket \; \le \; 2 \sqrt{2}
	\qquad \mbox{: quantum theory}
	\label{T quantum}
\end{equation}
of the expectation value for any state.

Next, by applying the trick (\ref{trick}) several times, we rewrite $ T $ as
\begin{eqnarray}
	T 
&=&
	2 \sqrt{2} \, (
	\sigma_x \otimes \sigma_x +
	\sigma_y \otimes \sigma_y +
	\sigma_z \otimes \sigma_z )
	\nonumber \\
&=&
	\frac{2 \sqrt{2}}{4} \Big[
	\sigma_x \otimes 
	\{ (\sigma_x + \sigma_y) + (\sigma_x - \sigma_y) 
	+  (\sigma_z + \sigma_x) - (\sigma_z - \sigma_x) \}
	\nonumber \\ && \qquad
	+ \sigma_y \otimes 
	\{ (\sigma_y + \sigma_z) + (\sigma_y - \sigma_z) 
	+  (\sigma_x + \sigma_y) - (\sigma_x - \sigma_y) \}
	\nonumber \\ && \qquad
	+ \sigma_z \otimes 
	\{ (\sigma_z + \sigma_x) + (\sigma_z - \sigma_x) 
	+  (\sigma_y + \sigma_z) - (\sigma_y - \sigma_z) \}
	\Big].
\end{eqnarray}
By introducing
\begin{eqnarray}
&&	A_1 = \sigma_x \otimes I, \qquad \qquad
	A_2 = \sigma_y \otimes I, \qquad \qquad
	A_3 = \sigma_z \otimes I, 
	\nonumber \\
&&	B_1 = \frac{1}{\sqrt{2}} \, I \otimes (\sigma_y + \sigma_z), \quad
	B_2 = \frac{1}{\sqrt{2}} \, I \otimes (\sigma_z + \sigma_x), \quad
	B_3 = \frac{1}{\sqrt{2}} \, I \otimes (\sigma_x + \sigma_y),
	\nonumber \\
&&	B_4 = \frac{1}{\sqrt{2}} \, I \otimes (\sigma_y - \sigma_z), \quad
	B_5 = \frac{1}{\sqrt{2}} \, I \otimes (\sigma_z - \sigma_x), \quad
	B_6 = \frac{1}{\sqrt{2}} \, I \otimes (\sigma_x - \sigma_y),  \quad
\end{eqnarray}
we reach the expression
\begin{eqnarray}
	T 
&=&
	A_1 ( B_3 + B_6 + B_2 - B_5 ) +
	A_2 ( B_1 + B_4 + B_3 - B_6 ) +
	A_3 ( B_2 + B_5 + B_1 - B_4 ) 
	\nonumber \\
&=&
	A_1 ( B_3 + B_6 ) 
	+ A_2 ( B_3 - B_6 ) 
	\nonumber \\ &&
	+ A_3 ( B_2 + B_5 ) 
	+ A_1 ( B_2 - B_5 ) 
	\nonumber \\ &&
	+ A_2 ( B_1 + B_4 ) 
	+ A_3 ( B_1 - B_4 ),
	\label{T}
\end{eqnarray}
which was shown at (\ref{T intro}) in Introduction.

The hidden variable theory is applicable to $ T $ in this form.
In the context of the hidden variable theory,
$ A_i $ and $ B_j $ are functions of the hidden variable $ \lambda $
and the values of $ A_i ( \lambda ) $ and $ B_j ( \lambda ) $ are 
in $ \{ 1, -1 \} $.
Then it is easily seen from the expression (\ref{T})
that the possible values of $ T ( \lambda ) $ are in $ \{ 6, 2, -2, -6 \} $.
Hence, the average 
$ \bra T \ket = \int T ( \lambda ) P ( \lambda ) d \lambda $
is in the range
\begin{equation}
	-6 \; \le \; \bra T \ket \; \le \; 6
	\qquad \mbox{: hidden variable theory}.
	\label{T hidden}
\end{equation}
The range (\ref{T quantum}) of the prediction of the quantum theory
and the range (\ref{T hidden}) of the hidden variable theory
have the overlap $ [ -6, 2 \sqrt{2} ] $ where both the theories hold,
and the region $ [ -6 \sqrt{2}, -6 ) \cup ( 2 \sqrt{2}, 6 ] $
where only one of the two theories holds.
Thus, $ T $ is a quantity which realizes the type 2 test.

Moreover, we introduce a parameter $ \theta $ and define $ T_\theta $
as the polynomial (\ref{T}) of
\begin{eqnarray}
&&	A_1 = ( \sigma_z \cos 2 \theta + \sigma_x \sin 2 \theta ) \otimes I, 
	\label{A1} \\
&&	A_2 = ( \sigma_z \cos 2 \theta + \sigma_y \sin 2 \theta ) \otimes I, \\
&&	A_3 = \sigma_z \otimes I, \\
&&	B_1 = I \otimes ( \sigma_y \cos \theta + \sigma_z \sin \theta ), \\
&&	B_2 = I \otimes ( \sigma_z \cos \theta + \sigma_x \sin \theta ), \\
&&	B_3 = I \otimes ( \sigma_x \cos \theta + \sigma_y \sin \theta ), \\
&&	B_4 = I \otimes ( \sigma_y \cos \theta - \sigma_z \sin \theta ), \\
&&	B_5 = I \otimes ( \sigma_z \cos \theta - \sigma_x \sin \theta ), \\
&&	B_6 = I \otimes ( \sigma_x \cos \theta - \sigma_y \sin \theta ).
	\label{B6}
\end{eqnarray}
The parameter $ \theta $ is interpreted as an angle 
which specifies the directions of the detectors. 
The matrix representation of $ T_\theta $ is
\begin{equation}
	T_\theta
	=
	2 (\cos \theta + \sin \theta ) 
	\begin{pmatrix}
	1 & (1-i) \cos 2 \theta & 0 & 0 \\
	(1+i) \cos 2 \theta & -1 & 2 \sin 2 \theta & 0 \\
	0 & 2 \sin 2 \theta & -1 & -(1-i) \cos 2 \theta \\
	0 & 0 & -(1+i) \cos2 \theta & 1 
	\end{pmatrix}.
\end{equation}
\begin{figure}[bt]
\begin{center}
\scalebox{0.50}{
\includegraphics{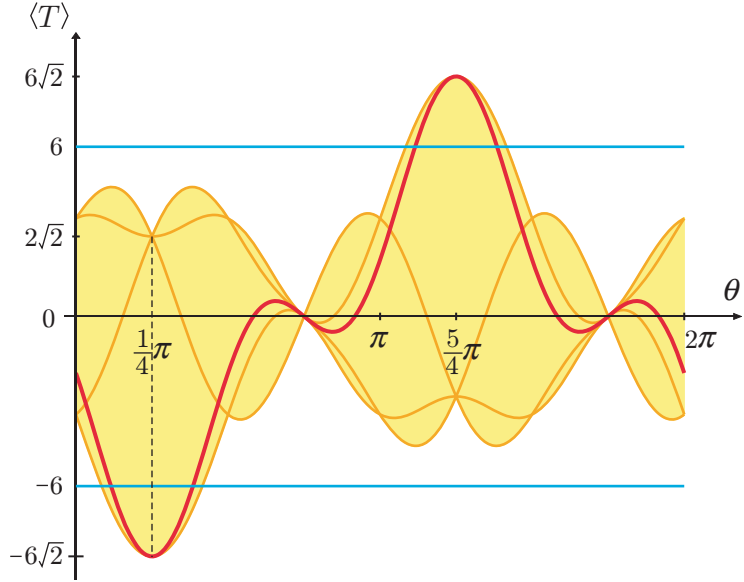}
}
\end{center}
\vspace{-2mm}
\caption{\label{FIG3}A test of 
the new inequality.
The hidden variable theory predicts that $ -6 \le \bra T \ket \le 6 $.
The quantum theory predicts that $ \bra T \ket $ is in the yellow band
$ B_T $ defined in (\ref{B_T}).
The red curve is the expectation value (\ref{singlet value for T})
associated with the singlet state.}
\end{figure}
When $ \theta = \frac{\pi}{4} $, 
$ T_\theta $ is reduced to the original form (\ref{starting T matrix}).
The eigenvalues of $ T_\theta $ are
\begin{eqnarray}
&&	\{ t_1 (\theta), 
	t_2 (\theta), 
	t_3 (\theta), 
	t_4 (\theta) \}
	\nonumber \\
&& =
	\Big\{
	2 ( \cos \theta + \sin \theta )
	\big(
		- \sin 2 \theta 
		\pm \sqrt{ \cos^2 2 \theta + 2 + 2 \sin 2 \theta}
	\big),
	\nonumber \\ && \quad \quad
	2 ( \cos \theta + \sin \theta )
	\big(
		\sin 2\theta 
		\pm \sqrt{ \cos^2 2 \theta + 2 - 2 \sin 2 \theta}
	\big)
	\Big\}.
\end{eqnarray}
The sets of values allowed by the quantum theory are denoted as
\begin{eqnarray}
	Q_T ( \theta ) 
	& := &
	\Big\{ 
	\sum_{i=1}^4 w_i t_i (\theta) \, \Big| \, 
	w_i \in \R, \, 0 \le w_i \le 1, \, \sum_{i=1}^4 w_i = 1 \,
	\Big\}
	\subset \R,
\\
	Q_T 
	& := & \bigcup_{ 0 \le \theta \le 2 \pi} Q_T ( \theta ) 
	= [ -6 \sqrt{2}, \; 6 \sqrt{2} \, ]
	\: \subset \R,
\\
	B_T 
	& := & \{ ( \theta, t ) \, | \, 
	0 \le \theta \le 2 \pi, \, t \in Q_T (\theta) \}
	\subset \R^2.
	\label{B_T}
\end{eqnarray}
And the range allowed by the hidden variable theory is denoted as
\begin{eqnarray}
	H_T := [ -6, \, 6 \, ] \: \subset \R.
\end{eqnarray}
In the figure \ref{FIG3} the band $ B_T $ is shown as the painted region.
At any fixed value of $ \theta $,
the predictions of the two theories have some overlap,
namely, we have $ Q_T ( \theta ) \cap H_T \ne \varnothing $.
It also happens that $ Q_T ( \theta ) \subset H_T $ 
for some $ \theta $.
But it never happens that $ Q_T ( \theta ) \supset H_T $ at any $ \theta $.
In this sense, the tests of type 2 and type 3 are realized.
It is also to be noted that $ Q_T \supset H_T $,
namely, the whole set of quantum predictions is wider than
predictions of the hidden variable theory.

In particular, the singlet state $ \psi $ of (\ref{singlet}) 
yields the expectation value
\begin{eqnarray}
	f ( \theta )
	= \bra \psi | T_\theta | \psi \ket
	= -2 ( \cos \theta + \sin \theta )
	( 1 + 2 \sin 2 \theta ).
	\label{singlet value for T}
\end{eqnarray}
The range 
$ F 
:= \{ f ( \theta ) \, | \, 0 \le \theta \le 2 \pi \} 
= [ -6 \sqrt{2}, \, 6 \sqrt{2} ] $
covers $ H_T $ completely.
This means that violation of the bound of the hidden variable theory
is observed with the state $ \psi $.

If we take another entangled state
\begin{eqnarray}
	\chi 
	=
	| \! \uparrow \uparrow \ket + | \! \downarrow \downarrow \ket 
	=
	\frac{1}{\sqrt{2}}
	\begin{pmatrix} 1 \\ 0 \end{pmatrix}
	\otimes
	\begin{pmatrix} 1 \\ 0 \end{pmatrix}
	+
	\frac{1}{\sqrt{2}}
	\begin{pmatrix} 0 \\ 1 \end{pmatrix}
	\otimes
	\begin{pmatrix} 0 \\ 1 \end{pmatrix}
	=
	\frac{1}{\sqrt{2}}
	\begin{pmatrix} 1 \\ 0 \\ 0 \\ 1 \end{pmatrix}
	\label{entangled}
\end{eqnarray}
instead of the singlet state $ \psi $, it yields
\begin{eqnarray}
	g ( \theta )
	= \bra \chi | T_\theta | \chi \ket
	= 2 ( \cos \theta + \sin \theta ).
\end{eqnarray}
The range 
$ G 
:= \{ g ( \theta ) \, | \, 0 \le \theta \le 2 \pi \} 
= [ -2 \sqrt{2}, \, 2 \sqrt{2} ] $
is included in $ H_T $ completely.
This means that violation of the hidden variable theory
will be never observed 
in measurement of $ T $ with the state $ \chi $.

\section{Discussions}
Here we give a summary of this study.
In this paper we pointed out that
violation of the BCHSH inequality can be understood as a result
of noncommutativity of quantum observables.
For noncommuting operators,
an eigenvalue of a sum of operators does not coincide with
a sum of eigenvalues of the respective operators.
Using this property,
we invented a method to build systematically  Bell-like observables
and showed that the conventional BCHSH inequality is reconstructed by this method.
This diminished the ad hoc nature of the BCHSH observable.

We classified possible tests of the hidden variable theory and the quantum theory.
We pointed out that
there was no chance in the conventional BCHSH test 
to reveal invalidity of the quantum theory 
with validity of the hidden variable theory.
By applying our method,
we constructed the new observable $ T $ and calculated the range of its average
in the contexts of the hidden variable theory and the quantum theory, respectively.
It was shown that
there is a chance that the new test with $ T $ reveals
invalidity of the quantum theory with validity of the hidden variable theory.

Of course,
we do not aim to deny validity of the quantum theory,
but we aim to support it
by passing the new severer test.
It is also our purpose to clarify
the implication of violation of various types of Bell-like inequalities. 

There are several remaining problems concerning the generalized inequality.
First one is the existence problem
in a mathematical sense.
The Bell-like inequality is a necessary condition for existence of 
the probability distribution which satisfies (\ref{local}).
For the case of the conventional BCHSH inequality,
Fine~\cite{Fine1982} proved that the set of inequalities
\begin{eqnarray}
&&	-2 \: \le \:
	  \bra A_1 B_1 \ket
	+ \bra A_1 B_2 \ket
	+ \bra A_2 B_1 \ket
	- \bra A_2 B_2 \ket
	\: \le \: 2,
	\\
&&	-2 \: \le \:
	  \bra A_1 B_2 \ket
	+ \bra A_2 B_1 \ket
	+ \bra A_2 B_2 \ket
	- \bra A_1 B_1 \ket
	\: \le \: 2,
	\\
&&	-2 \: \le \:
	  \bra A_2 B_1 \ket
	+ \bra A_2 B_2 \ket
	+ \bra A_1 B_1 \ket
	- \bra A_1 B_2 \ket
	\: \le \: 2,
	\\
&&	-2 \: \le \:
	  \bra A_2 B_2 \ket
	+ \bra A_1 B_1 \ket
	+ \bra A_1 B_2 \ket
	- \bra A_2 B_1 \ket
	\: \le \: 2
\end{eqnarray}
is a necessary and sufficient condition for existence of
the probability distribution of the hidden variable.
Although there are studies on tightness of some variations 
of the BCHSH inequalities~\cite{Garg1984, Laskowski2004, Pal2009},
finding the necessary and sufficient condition for existence of
the probability distribution of the hidden variable
for our observable $ T $ is left as an open problem.

The second one is a practical problem.
In principle, the test which we proposed can be implemented in experiment
using pairs of photons or spin-half particles.
But our choice involves nine observables (\ref{A1})-(\ref{B6})
with variable angle.
Hence, our scheme is still too cumbersome for practical use.
It is more desirable to reduce the number of observables
to make experiments easier.

The third one is extensibility of our scheme.
The proposed observable $ T $ is defined in the two-qubit Hilbert space.
It is possible to extend our scheme to multi-qubit systems.
It is also desirable to construct a Bell-like quantity with less number of observables.

\section*{Acknowledgements}
We would like to express our sincere thanks
to Prof. T.~Iwai and Prof. Y.~Y.~Yamaguchi 
for their valuable comments on our study.
We thank Prof. Valerio Scarani, who taught us 
the work by A.~Fine
and the work by D.~Collins and N.~Gisin;
both are related to tightness of the BCHSH inequality.
The referee also gave us comments useful for improving our manuscript.
We thank 
I.~Tsutsui,
T.~Ichikawa,
M.~Ozawa,
A.~Hosoya,
M.~Kitano,
H.~Kobayashi, and
S.~Ogawa
for their interests in our study and for their helpful comments.
This work was supported 
by the Grant-in-Aid for Scientific Research
of Japan Society for the Promotion of Science,
Grant No.~22540410.

%

\end{document}